\documentclass[aps,prl,preprint,showpacs,groupedaddress]{revtex4}
\usepackage {amssymb}
\usepackage {amsmath}
\usepackage {graphicx}
\usepackage {longtable}

\begin{document}
\title{High - temperature phase transition in a plasma and the mechanism
of powerful solar flares}

\author{Fedor V.Prigara}
\affiliation{Institute of Microelectronics and Informatics,
Russian Academy of Sciences,\\ 21 Universitetskaya, Yaroslavl
150007, Russia} \email{fprigara@imras.net76.ru}

\date{\today}

\begin{abstract}

It is shown that the high- temperature phase transition in a plasma gives
the mechanism of transition from the highly conductive state to the highly
resistive state of a plasma in the `electric circuit' model of solar flares
which was first introduced by H.Alfven and P.Carlqvist in 1967. With this
addendum, the modern version of the electric circuit model can explain both
the fast dissipation of energy and the acceleration of particles in a solar
flare.

\end{abstract}


\maketitle

Among various models of solar flares, the `electric circuit' model
[1,2], which was first introduced by Alfven and Carlqvist in 1967,
seems to be the most adequate for explaining of the fast
dissipation of energy and the acceleration of particles. The main
problem of the electric circuit model was so far the mechanism of
transition from the highly conductive state to the highly
resistive state of a plasma in a magnetic loop [1,2]. It was shown
recently [3] that a hot plasma can undergo the phase transition
from the cold phase to the hot phase. The transport properties of
the hot phase essentially differ from those of the cold phase due
to the intense interaction of the plasma with thermal radiation
above the inversion temperature [4]. The specific resistivity of
the hot phase, $\eta _{h} $, is about three orders of magnitude
larger than that of the cold phase, $\eta _{c} $, so the high-
temperature phase transition in a plasma can supply a mechanism
triggering the fast increase in the electric resistivity of a
plasma column in a magnetic loop. Below we consider this mechanism
in more detail.

The critical value $T_{e,c} $ of the electron temperature,
corresponding to the transition from the cold to the hot phase,
for relatively low plasma densities achievable in tokamaks is
independent of the density and has an order of magnitude of the
inversion temperature $T_{0} \cong 2keV$ [4]. Studies of sawtooth
oscillations in tokamak discharges by von Goeler, Stodiek, and
Southoff [5] suggest that the transition temperature is
approximately $T_{e,c} \cong 0.8keV$. The line of phase
equilibrium between the hot and cold phases of a plasma is
described by the following approximate expressions:

\begin{equation}
\label{eq1}
T_{c} \cong T_{c,0} ,
\quad
n < n_{c,0} ,
\end{equation}

\begin{equation}
\label{eq2}
n_{c} \cong 12T/\left( {hc\sigma _{a}}  \right),
\quad
n > n_{c,0} ,
\end{equation}

\noindent
where $n_{c,0} \cong 0.8 \times 10^{17}cm^{ - 3}$ for $T_{c,0} \cong
0.8keV$, \textit{c} is the speed of light, \textit{h} is the Planck
constant, \textit{T} is the temperature, the Boltzmann constant being
assumed to be included in the definition of the temperature, \textit{n} is
the plasma density, and$\sigma _{a} $ is the absorption cross-section.

The specific resistivity of a hot plasma has a form [6]

\begin{equation}
\label{eq3}
\eta = m_{e} \nu _{ei} /\left( {ne^{2}} \right),
\end{equation}

\noindent where the frequency of electron-ion collisions in the
cold phase of a plasma is given by the formula [6,7]

\begin{equation}
\label{eq4}
\nu _{ei} = nv_{Te} \sigma _{ei} \approx 2\pi e^{4}nlog\left( {n\lambda
_{D}^{3}}  \right)/\left( {m_{e}^{2} v_{Te}^{3}}  \right),
\end{equation}

\noindent
where $n = n_{e} = n_{i} $ is the electron and ion density, $v_{Te} $ is the
thermal velocity of electrons, $\sigma _{ei} $ is the electron-ion
collisional cross-section, $m_{e} $and $e$ are the mass and charge of
electron respectively, and $\lambda _{D} $ is the Debye length. It is clear
from equation (\ref{eq4}) that the Coulomb collisional cross-section is very small
at high temperatures, $\sigma _{ei} \propto T_{e}^{ - 2} $, where $T_{e} $
is the electron temperature.

It has been shown recently [8] that the frequency of electron-ion
collisions in the hot phase of a plasma interacting with thermal
radiation is given by the formula

\begin{equation}
\label{eq5}
\nu _{ei} = nv_{Te} \sigma _{ei} \approx nv_{Te} \sigma _{0} ,
\end{equation}

\noindent where $n = n_{e} = n_{i} $ is the plasma density,
$v_{Te} $ is the thermal velocity of electrons, and $\sigma _{ei}
$ is the cross-section of electron-ion collisions which is a
constant, $\sigma _{0} $, determined by the atomic size. The
origin of the constant cross-section $\sigma _{0} $ is as follows.
A hot plasma with the temperature $T_{e} \geqslant T_{0} \cong
2keV$ is intensely interacting with the field of thermal
radiation. At temperatures $T \geqslant T_{0} $ the stimulated
radiation processes dominate this interaction [9]. Thermal
radiation induces radiative transitions in the system of electron
and ion which corresponds to the transition of electron from the
free to the bound state.

Thus, in a hot plasma interacting with thermal radiation, the bound states
of electrons and ions restore, leading to the change of collisional
properties of a hot plasma. In this case, the electron-ion collisional
cross-section has an order of magnitude of the atomic cross-section, $\sigma
_{0} \cong 10^{ - 15}cm^{2}$.

Since the specific resistivity $\eta $of a plasma is proportional to the
frequency of electron-ion collisions, as given by equation (\ref{eq3}), the
resistivity of a plasma in the hot phase increases with the electron
temperature as

\begin{equation}
\label{eq6}
\eta _{H} \propto T_{e}^{1/2} ,
\end{equation}

\noindent
contrary to the relation for the cold phase of a plasma

\begin{equation}
\label{eq7}
\eta _{C} \propto T_{e}^{ - 3/2} .
\end{equation}

From equations (\ref{eq3}), (\ref{eq4}), and (\ref{eq5}) we find the ratio of the specific
resistivities for the hot and cold phases of a plasma respectively at the
transition temperature to be approximately $\eta _{H} /\eta _{C} \cong 2
\times 10^{3}$.

According to equations (\ref{eq3}) and (\ref{eq4}), the specific
resistivity of the cold phase nearby the transition temperature is
$\eta _{C} \cong 10^{ - 5}\Omega m$. If the length of a corona
loop is about $l \cong 10^{10}cm$, and its cross- section is $S
\cong 10^{18}cm^{2}$[2], then the resistivity of the magnetic loop
is $R \cong 10^{ - 11}\Omega $. The mean electric current through
a magnetic loop is about $I \cong 3 \times 10^{11}A$, so the mean
rate of energy dissipation in the cold phase in the vicinity of
the transition temperature is

\begin{equation}
\label{eq8}
dW/dt = I^{2}R \cong 10^{19}ergs^{ - 1}.
\end{equation}

This value is some nine orders of magnitudes less than the energy
dissipation rate in powerful solar flares [2].

There are two mechanisms of the enlarging resistivity of a
magnetic loop. One is the phase transition to the hot phase of a
plasma which gives an increase of resistivity $R_{H} /R_{C} \cong
2 \times 10^{3}$, as indicated above. The temperatures of the
plasma in solar flares inferred from the observational data are
about $5keV$ [2], what exceeds the critical temperature of the
high-temperature phase transition in a plasma. Another mechanism
is the radial contraction of the current channel in a magnetic
loop due to poloidal (transverse) magnetic field generated by the
current. If the diameter of the current channel changes from the
initial value of $d_{0} \cong 10^{9}cm$ to the final value of $d
\cong 10^{7}cm$, then the increase of the resistivity of the
magnetic loop is $R/R_{0} = S/S_{0} = d^{2}/d_{0}^{2} \cong
10^{4}$. These two mechanisms of the growth of the magnetic loop
resistivity are sufficient since the rate of energy dissipation in
powerful solar flares was overestimated in the models based on the
synchrotron emission generated by fast electrons (see below).

In fact, the contraction of the current channel leads to its disintegration
into the separate filaments, so that the total cross- section of the
magnetic loop remains nearly constant, while the sum of cross- sections for
the current filaments decreases essentially with respect to the initial
cross- section of the current channel. Thus, the sequence of events during a
powerful solar flare is as follows: (i) the contraction of the current
channel in a coronal magnetic loop; (ii) the disintegration of the current
channel into the separate filaments; (iii) the high- temperature phase
transition in the plasma of the current filaments.

The value of the poloidal magnetic field generated by the current in a
magnetic loop is given by the formula

\begin{equation}
\label{eq9}
B_{p} = 2I/\left( {cr} \right),
\end{equation}

\noindent where \textit{r} is the radius of the current channel.
The value of the poloidal magnetic field is $B_{p0} \cong 60G$ for
$r_{0} \cong 10^{9}cm$, and $B_{p} \cong 6 \times 10^{3}G$ for $r
\cong 10^{7}cm$. The mean toroidal (longitudinal) magnetic field
in the loop is about $B_{T} \cong 400G$ [2]. The safety factor
defined by the formula

\begin{equation}
\label{eq10}
q = B_{T} r/\left( {B_{p} R} \right),
\end{equation}

\noindent where \textit{R} is the radius of the magnetic loop, is
small, so the Kruskal- Shafranov condition for kink instability is
satisfied [10]. It means that the current does not flow along the
magnetic loop axis, but instead a current line is helical.

The mechanism of the electric field generation in a magnetic loop
is described in Ref.2. The accelerating electric field can be
roughly estimated as

\begin{equation}
\label{eq11}
E \cong vB/c,
\end{equation}

\noindent where \textit{v} is the velocity of convection motions,
$v = 0.1 - 0.5kms^{ - 1}$ [2]. If the value of the magnetic field
is $B \cong 10^{3}G$, then the accelerating electric field is $E
\cong 0.3Vcm^{ - 1}$. For the characteristic scale of the magnetic
loop, $l \cong 10^{9}cm$, the maximum energy of accelerated
particles is $W_{im} = eEl \cong 300MeV$, what is comparable with
the maximum energy of fast ions in a powerful solar flare [2]. The
maximum energy of accelerated electrons is restricted by their
mean free path with respect to the electron- ion collisions,

\begin{equation}
\label{eq12}
l_{e} = v_{Te} /\nu _{ei} = 1/\left( {n\sigma _{0}}  \right)
\end{equation}

\noindent in the hot phase of the plasma. For the plasma density
$n \cong 10^{10}cm^{ - 3}$, the mean free path of electrons is
$l_{e} \cong 10^{5}cm$, and the maximum energy of electrons is
$W_{em} \cong eEl_{e} \cong 30keV$. This value is consistent with
the available data on the electron energy distribution in powerful
solar flares [2].

The disruptive instability [11] of the current channel in the hot
phase of the plasma can be responsible for the ejection of the
plasma with accelerated particles in the external space (see also
Ref.8).

Due to the Colgate paradox [2], the X- ray emission from solar
flares cannot have a synchrotron origin. Presumably, it is
produced by the reverse Compton scattering modified by the
stimulated radiation processes in the hot phase of the plasma
interacting with thermal radiation.

To summerize, we show that the account for the high- temperature
phase transition in a plasma can essentially improve the
possibility of the 'electric circuit' model to explain the
phenomena observed in powerful solar flares.

\begin{center}
---------------------------------------------------------------
\end{center}

[1] H.Alfven and P.Carlqvist, Sol. Phys. \textbf{1}, 220 (1967).

[2] V.V.Zaitsev and A.V.Stepanov, Usp. Fiz. Nauk \textbf{176}, 325
(2006) [Physics- Uspekhi \textbf{49} (2006)].

[3] F.V.Prigara, E-print archives, physics/0509199.

[4] F.V.Prigara, E-print archives, physics/0503197.

[5] S.von Goeler, W.Stodiek, and N.Sauthoff, Phys. Rev. Lett.
\textbf{33}, 1201 (1974).

[6] F.F.Chen, \textit{Introduction to Plasma Physics and
Controlled Fusion, Vol.1: Plasma Physics} (Plenum Press, New York,
1984).

[7] V.P.Silin, Usp. Fiz. Nauk \textbf{172}, 1021 (2002), [Physics-
Uspekhi \textbf{45}, 955 (2002)].

[8] F.V.Prigara, E-print archives, physics/0410102.

[9] F.V.Prigara, E-print archives, physics/0404087.

[10] S.C.Hsu and P.M.Bellan, Phys. Plasmas \textbf{12}, 032103
(2005).

[11] B.B.Kadomtsev, \textit{Collective phenomena in plasmas}
(Nauka, Moscow, 1988).

\end{document}